\documentclass[fleqn,usenatbib]{mnras}

\usepackage{newtxtext,newtxmath}

\usepackage[T1]{fontenc}

\DeclareRobustCommand{\VAN}[3]{#2}
\let\VANthebibliography\thebibliography
\def\thebibliography{\DeclareRobustCommand{\VAN}[3]{##3}\VANthebibliography}

\usepackage{graphicx}   
\usepackage{amsmath}    

\title[A nebula associated with a fast sdB star]{Discovery of a 
nebula associated with a high proper motion sdB star} 

\author[R.\ Ortiz et al.]{
R. Ortiz,$^{1}$\thanks{E-mail: rortiz@usp.br}
F. Bijarchian,$^{2}$
M.A. Guerrero,$^{2}$
M. Akhlaghi,$^{3,4}$
and M. Serra-Ricart $^{5,6,7}$
\\
$^{1}$Escola de Artes, Ci\^encias e Humanidades, USP, Av. Arlindo Bettio 1000, 03828-000 S\~ao Paulo, Brazil \\
$^{2}$Instituto de Astrof\'{i}sica de Andaluc\'{i}a, IAA-CSIC, Glorieta de la Astronom\'{i}a S/N, Granada 18008, Spain \\
$^{3}$Centro de Estudios de Física del Cosmos de Aragón (CEFCA), Plaza San Juan 1, 44001 Teruel, Spain \\
$^{4}$Unidad Asociada CEFCA-IAA, CEFCA, Unidad Asociada al CSIC por el IAA y el IFCA, Plaza San Juan 1, 44001 Teruel, Spain \\
$^{5}$Instituto de Astrofísica de Canarias (IAC), Vía Láctea, s/n, E-38205, La Laguna, Tenerife, Spain \\
$^{6}$Departamento de Astrofísica, Universidad de La Laguna (ULL), E-38206 La Laguna, Tenerife, Spain \\
$^{7}$Light Bridges, Observatorio Astronómico del Teide. Carretera del Observatorio del Teide, s/n, E-38570 Güímar, Tenerife, Spain 
}

\date{Accepted XXX. Received YYY; in original form ZZZ}

\pubyear{2026}

\begin{document}
\label{firstpage}
\pagerange{\pageref{firstpage}--\pageref{lastpage}}
\maketitle

\begin{abstract}
All B-type subdwarf stars (hereafter sdB) should have low flux of ionizing photons, making them incapable of producing a noticeable circumstellar photoionized shell. 
However, a few sdB stars have been associated with circumstellar nebulae, resembling in some cases a planetary nebula. 
These discoveries spark doubts about the nature of the physical processes behind the formation of the nebula. 
In this paper, we describe the newfound parabolic-shaped nebula associated with the high proper motion sdB star TYC\,3315-1807-1. 
The apex of the H$\alpha$ nebula is situated $\simeq0.5$ arcmin in the direction of the stellar proper motion. 
A wider parabolic-shaped nebula is also detected in {\it WISE} W1 infrared images at 3.4 $\mu$m, whereas {\it GALEX} images show extended far-UV emission around the star within the optical and mid-IR emissions. Like most other sdB stars with associated nebulae, TYC\,3315-1807-1 moves at a high-speed (102 km s$^{-1}$) across the Galactic plane. 
The low luminosity of TYC\,3315-1807-1 cannot provide its wind with the momentum necessary to form and keep a bow shock. 
The nebula around TYC\,3315-1807-1 is rather suggested to be a Mach wave partially excited by shocks and photoionization or the encounter of the star with an over-density clump in the ISM. 
\end{abstract}




\section{Introduction} 


Hot subdwarfs (hereafter `sd') are low-mass ($\lesssim 0.5 M_{\odot}$) evolved stars on the blue side of the horizontal branch (HB). 
They can be classified as sdB, which are core helium-burning stars, or sdO, which can be post-RGB, post-HB or post-AGB stars \citep{Heber2009}. 
Consequently, sdBs are spectroscopically homogeneous, generally showing abnormally broad Balmer lines and weak He lines, whereas sdOs exhibit a wider spectral variety, showing strong Balmer (sdO type) or He lines (those classified as He-sdO), as well as C or N strong lines \citep{Stroer2007}. 

Post-AGB sdO stars are in an evolutionary phase between the ejection of a planetary nebula (PN) and the formation of a white dwarf.  
These are expected to be surrounded by remnants of their ejected PNe, which would help to distinguish them from post-RGB and post-HB sdOs. However, extensive searches have produced a limited sample of 19 sdOs surrounded by nebulae, most of them associated with evolved, low-surface brightness PNe of complex morphologies around a binary central star \citep{Aller2015,Aller+2015}.

Meanwhile, compared to sdO stars, sdBs are less likely to exhibit ionized nebulae, as their UV flux is generally too low to ionize them.
Still, a few nebulae have been associated with sdB stars
\citep[Table A.1 in][]{G-S2021}, but some of these detections still lack confirmation (Table \ref{tab:sdBPN}). These might be associated with binaries, as in the case of nebulae around sdOs, since a volume-limited sample has shown that $\approx 30$\% sdBs in the solar neighborhood belong to binary systems \citep{SW2003}. 
Almost half of them have orbital periods shorter than 10 days \citep{Maxted2001,Napiwotzki2004}. 
Those with a near-IR excess must have a low-luminosity, cool companion.

The general availability of small to medium-sized telescopes with a wide field-of-view (FoV) and H$\alpha$ narrow-band filters allows sensitive searches for low surface-brightness nebulae.
Following similar searches for other nebular sources \citep[e.g., nova shells around CVs,][]{Sahman+2015}, we have started a program devoted to obtain deep and large FoV images using 1.0-m to 1.5-m aperture telescopes to search for nebular emission around sdB stars that seem to show extended emission in multi-wavelength observations.

This paper presents the first results of this program, reporting the detection of a rare case of a nebula associated with TYC\,3315-1807-1 (hereafter TYC\,3315, RA=03:21:39.63, Dec=+47:27:18.8, J2000) {\it aka} Cl\,Melotte\,20\,488. 
TYC\,3315 is an sdB star formerly identified by its near-UV to optical emission excess \citep{Kawka2010} that belongs to a binary system with an orbital period of $0.26584\pm0.00004$ day determined from radial velocity variations and its {\it NSVS} light curve \citep[{\it Northern Sky Variability Survey},][]{Wozniak2004}. 
A fit to the line profiles using a grid of appropriate stellar models implies the following stellar parameters: $T_{\rm eff}= 29,200\pm300$ K, $\log g = +5.5\pm0.1$, and $\log (n_{\rm He}/n_{\rm H})=-2.6\pm0.1$ \citep{Kawka2010}.  
Using these values, a fit to the observed broad-band near-UV to near-IR spectral energy distribution (SED) requires a colour-excess $E(B-V)=0.23$ \citep{Kawka2010}. 
That work assumed a stellar mass of $0.47~\mathrm{M}_{\odot}$, which was later reduced to $0.27~\mathrm{M}_{\odot}$ by \citet{Devarapalli2022}.  
Both works agree that the cool companion is a low-mass star with $M\simeq0.11-0.13~\mathrm{M}_{\odot}$ and dM5 (or earlier) spectral type, but neither report the presence of emission lines or peculiar features in their high- and low-resolution spectra. 

Interestingly, TYC\,3315 is a high proper motion star (Gaia DR3 435211617384833536), with $\mu_\alpha = 58.716\pm0.039$ mas~yr$^{-1}$ and $\mu_\delta = -8.208\pm0.034$ mas~yr$^{-1}$.
At the Gaia distance of $263.1\pm2.4$ pc these imply
a tangential velocity of $75.6 \pm 1.1$ km~s$^{-1}$. 
Together with its systemic velocity \citep[70.5 km~s$^{-1}$,][]{Kawka2010}, we obtain the space velocity $V_{\star}=102.1$ km~s$^{-1}$ and the Galactic systemic velocity $(U,V,W)=(-98.5,-16.5,+21.3)$ km s$^{-1}$. The inclination of its systemic velocity relative to the plane of the sky is $\approx 43^\circ$. 

The next sections report the observations and data reduction (Sect.\,2), main results (Sect.\,3), discussion (Sect.\,4), and concluding remarks (Sect.\,5). 

\begin{figure*}
\includegraphics[width=1.0\textwidth]{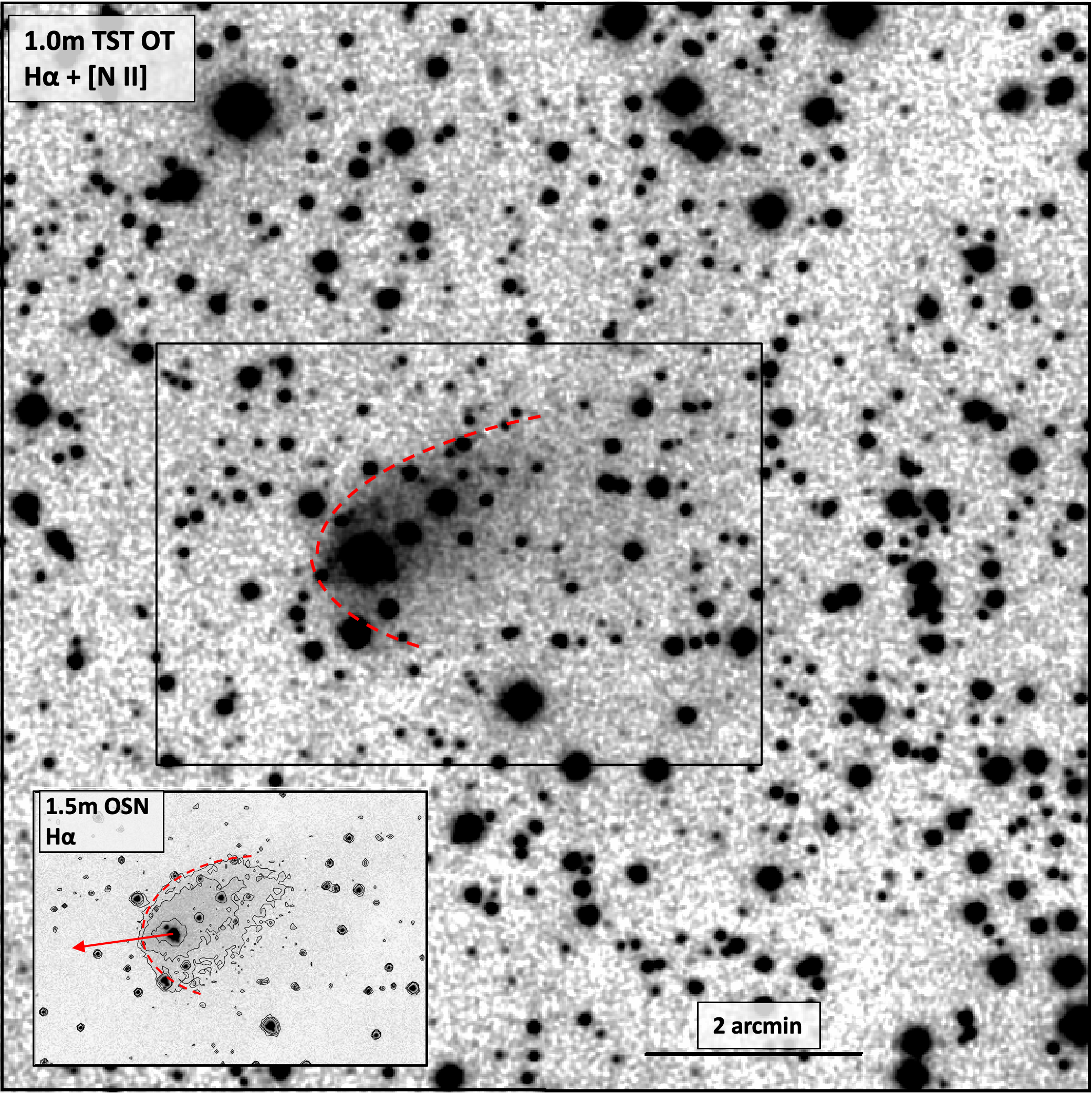}
\caption{
TST $10^\prime\times10^\prime$ H$\alpha$+[N~{\sc ii}] image of TYC\,3315. 
The inset ($5.7^\prime\times4.0^\prime$) shows the OSN 1.5m telescope H$\alpha$ image with contours at 1-, 2-, and 3-$\sigma$ over the background.  
The nebula outline, which follows the outermost contours, is highlighted by the same parabolic-shaped red-dashed curve in both images. The red arrow in the inset shows the stellar motion on the plane of the sky in a time-lapse of 1000 yr. North is up, east to the left.}
\label{fig:img}
\end{figure*}

\section{Observations and Data Reduction} \label{sec:obs}

TYC\,3315 was imaged with the 1.5-m telescope at Observatorio de Sierra Nevada ({\it OSN}, Granada, Spain) on January 9, 2025 using the CCDT150 camera, a $2048\times2048$ Andor ikon-L CCD.  
An on-chip binning $2\times2$ was applied, resulting in a pixel scale of 0.464 arcsec~pix$^{-1}$ and FoV $\simeq$8 arcmin.  
Ten 1200 s exposures were obtained with the OSN H01 H$\alpha$ narrow-band filter ($\lambda_\mathrm{c} = 6565$ \AA, $\Delta\lambda = 13$ \AA) with a 5-point dithering pattern with a throw of 90 arcsec.  
The images were bias subtracted, flat-fielded using suitable sky flat-field exposures, aligned, and then combined.  
The spatial resolution derived from the FWHM of stars in the FoV is 2.0 arcsec. 

To further check for the presence of emission at large spatial scales, TYC\,3315 was also imaged with the 1.0-m Transient Survey Telescope (TST) at Observatorio del Teide ({\it OT}, Tenerife, Spain) on February 19, 2025. 
The prime focus camera FERVOR-L (\textit{Fast Embedded-sCMOS Robotic Visible Observatory for Rapid transients}), based on sCMOS sensor (Sony IMX411, $14304\times10748$), was used.
FERVOR-L has a pixel scale of 0.60 arcsec~pix$^{-1}$ and an impressive FoV $2.4\times1.8$ degree$^2$. 
Fifty-seven 180 s exposures were obtained with an H$\alpha$+[N~{\sc ii}] filter ($\lambda_\mathrm{c} = 6572$ \AA, $\Delta\lambda = 64$ \AA) with dithering between exposures of 30 arcmin to minimize possible spurious background structures.  
The images were bias-subtracted and flat-fielded using sky flat–field exposures. 
Source detection and preliminary cleaning were performed with {\it NoiseChisel} \citep{gnuastro2015, gnuastroSegment2019a}, a non-parametric, noise-based detection algorithm implemented in GNU Astronomy Utilities \citet[\textit{Gnuastro},][]{gnuastro2025}.
The final spatial resolution, estimated from the FWHM of stars across the FoV, is 2.5 arcsec.

For those stars within the FoV of the TST image we used Gaia XP spectra as reference standards. To derive the reference magnitudes, we developed a script\footnote{https://codeberg.org/Gnuastro/scripts/src/branch/master/gaia-spectra-to-mag.sh} that computes synthetic photometry from Gaia XP spectra and a given transmission curve. The script multiplies  each spectrum with the transmission curve of the TST filter to obtain a band-averaged flux \citep{Eskandarlou2023,Gaia2023}, which is then converted to AB magnitudes. These synthetic magnitudes were compared with the instrumental magnitudes measured in the image to determine the photometric zeropoint for each exposure. The effective wavelength ($\lambda_\mathrm{c} = 6576.4$ \AA) and effective width ($\Delta\lambda =  68.22$ \AA) of the filter, computed using the filter transmission curve and a Vega spectrum, were used for this task. 
These zero-points were then homogenized to a common reference value, 
which enabled consistent scaling of the frames and facilitated outlier rejection through a defined sigma-clipping threshold.
The conversion factor from counts-per-second to physical flux (in units of erg~cm$^{-2}$~s$^{-1}$) was computed using the filter transmission at 6563 \AA, the wavelength of the H$\alpha$ emission line.  
The scaled frames were then aligned and combined using the {\it Gnuastro} {\it Warp} \citep[Sect. 6.4.4 of][]{gnuastro2025} and {\it Arithmetic} \citep[Sect. 6.2.4.7 of][]{gnuastro2025} programs, respectively, to produce a final coadded flux-calibrated image.  
Note that the pixel scale of the final coadded image is smaller than that of the original images in order to eliminate the Moir\'e pattern\footnote{The Moir\'e pattern is a noticeable non-flat noise that occurs when two slightly different grids are super-imposed.  
It can be corrected using a smaller pixel scale on the output images.} \citep[see Sect.\ 2.9 of][]{gnuastro2025}.

\begin{figure*}
\includegraphics[width=0.32\textwidth]{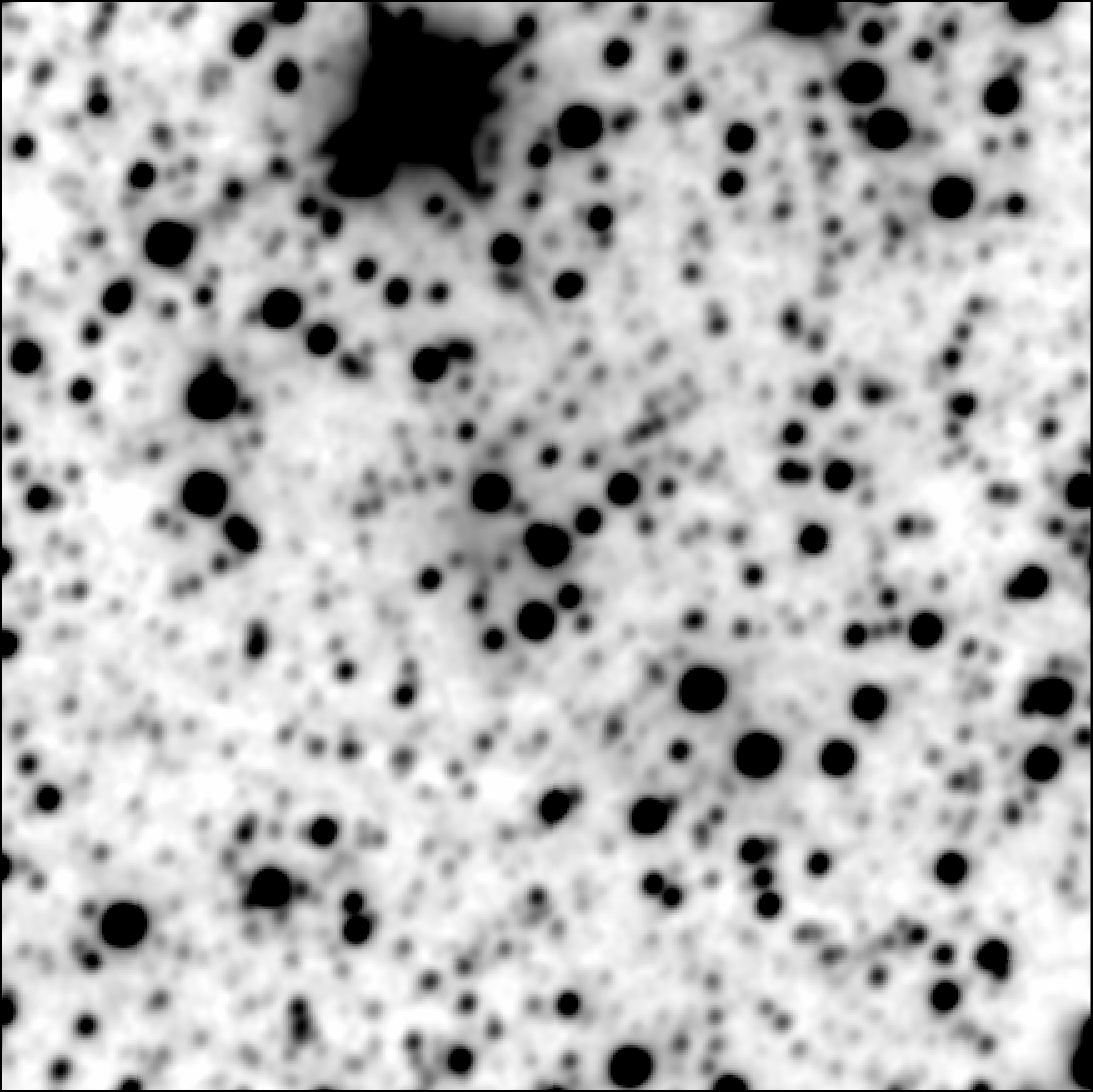}
\includegraphics[width=0.32\textwidth]{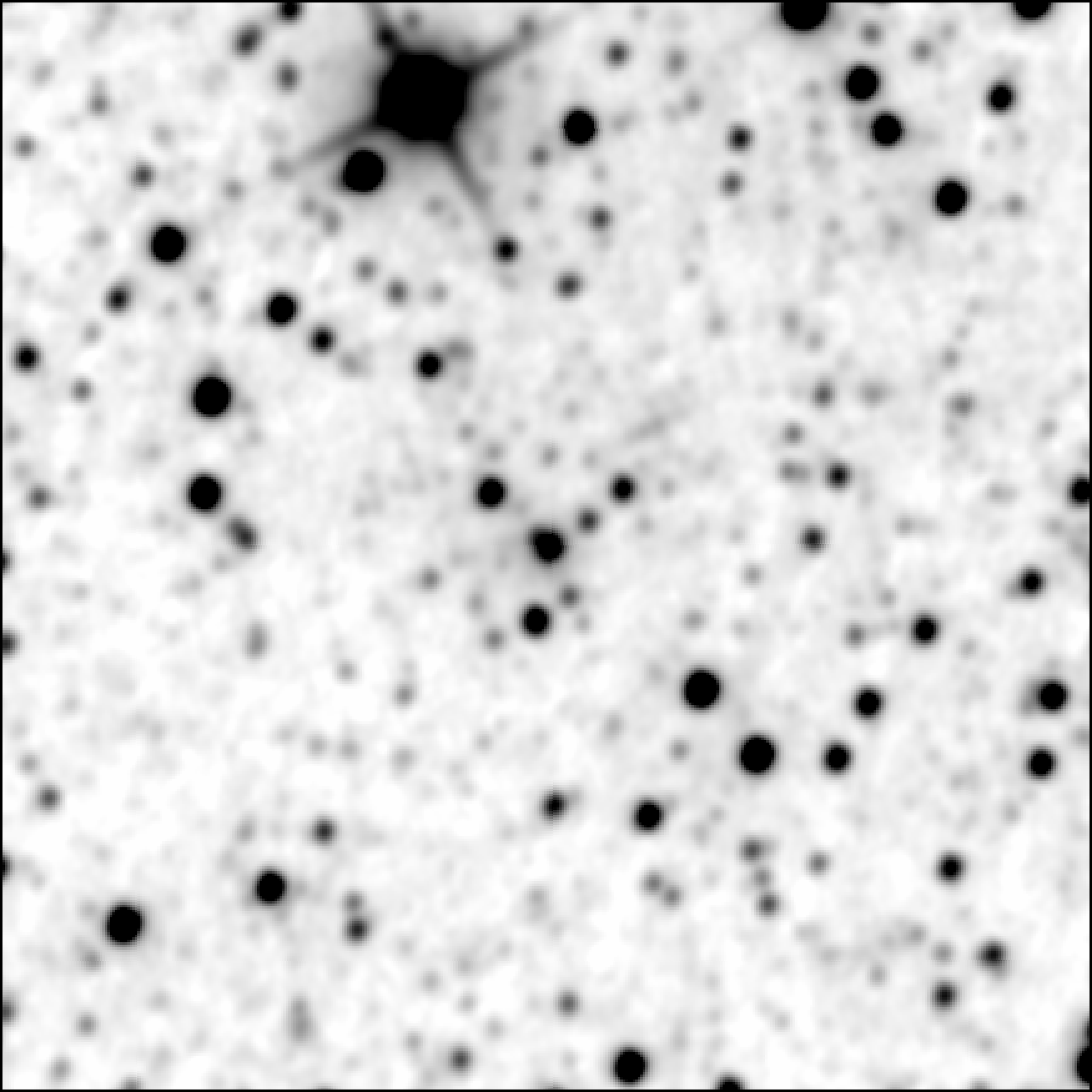}
\includegraphics[width=0.32\textwidth]{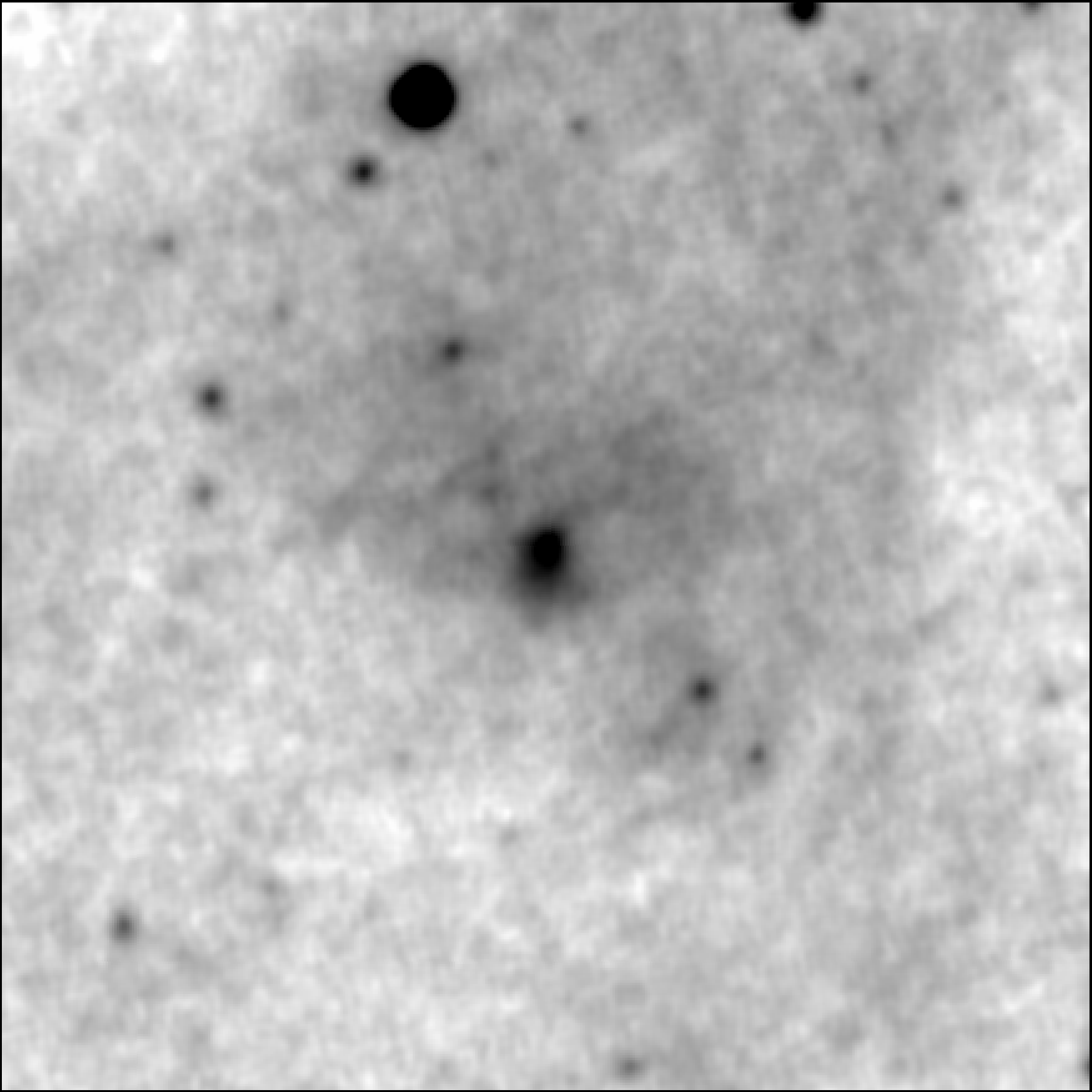}
\caption{
{\it WISE} W1 3.4 $\mu$m (left), W2 4.6 $\mu$m (centre), and W3 12 $\mu$m (right) inverted grey-scale images of TYC\,3315. The FoV is $10^\prime\times10^\prime$.  
North is up, east to the left.}
\label{wise}
\end{figure*}

\section{Results} \label{sec:result}

\subsection{Detection of diffuse H$\alpha$ emission}
\label{halpha}

The {\it OSN} 1.5-m telescope and TST images of TYC\,3315 reveal the presence of an extended parabolic-shaped nebula (Fig.~\ref{fig:img}).  
Its apex is located at $\approx 28$ arcsec from the star along a position angle (PA) $\approx 92^{\circ}$.  
The existence of this feature can be confirmed by examining the high-contrast TST H$\alpha$+[N~{\sc ii}] image shown in the same Fig.~\ref{fig:img}.
This parabolic-shaped nebula has a height of $\approx $3.5 arcmin and its width at the base is $\approx $2.7 arcmin.
It does not show a clear limb-brightening, but its surface brightness is rather smooth, suggesting a large filling factor. 
The TST image also reveals low-level variations of the sky surface brightness on large-spatial scales across the field of view.

To determine the H$\alpha$ flux of this nebula, a source aperture encompassing it was defined. To isolate the nebular emission within the aperture, contaminating sources were first detected using {\it Gnuastro}’s Segment \citep{gnuastroSegment2019a} and {\it MakeCatalog} \citep{makecatalog2019b} programs, and then masked using {\it MakeProfile} \citep[Sect. 8.1.4 of][]{gnuastro2025}.
Suitable background regions with comparable stellar density near the source aperture were selected to estimate the local background level in a similar way, which was subsequently subtracted from the science region. 
Within the chosen source aperture, the net integrated nebular flux is 
$6.4 \times 10^{-13}$ erg~cm$^{-2}$~s$^{-1}$. 
The major source of uncertainty is not the flux calibration, but the removal of the background level, as the emission of this background is noted to vary across the field of view. 
A series of tests performed selecting different background regions indicates that the flux uncertainty of the bow-shock is less than 15\%. The intrinsic H$\alpha$ flux is computed correcting it for $E(B-V)= 0.23$ \citep{Kawka2010}.
This reddening implies a correction factor of 1.4, resulting in an intrinsic H$\alpha$ flux of (9.0 $\pm$ 1.3) $\times 10^{-13}$ erg cm$^{-2}$ s$^{-1}$.

The mass of the ionized nebula $M_\mathrm{neb}$ around TYC\,3315 can be derived from its intrinsic H$\alpha$ flux as \citep[see e.g.][]{Boyarchuk+1968} 

\begin{equation}
    M_\mathrm{neb} = \mu m_{\rm p} \sqrt{\frac{4 \pi d^2 \varepsilon V F_\mathrm{H\alpha}}{j_\mathrm{H\alpha}}},
\end{equation}

\noindent where $\mu$ is the mean molecular weight, m$_{\mathrm p}$ is the proton mass, $d$ is the distance, $\varepsilon$ is the filling factor, $V$ is the emitting volume, $F_{{\mathrm H}\alpha}$ is the unabsorbed (intrinsic) H$\alpha$ flux, and 
$j_\mathrm{H\alpha} = 4\times10^{-25}$ erg~cm$^{-3}$~s$^{-1}$ \citep{OF2006} is the H$\alpha$ line emission coefficient. 
We assume $\mu = 1.44$, corresponding to the solar abundance ratio of the ISM. Most of the emission from the nebula can be inscribed within a paraboloid with a circular base of radius 1.4 arcmin and height of 3.5 arcmin, which corresponds to a deprojected height of 4.8 arcmin, assuming an inclination relative to the plane of the sky of $\approx 43^\circ$. This implies a volume of the emitting nebula of $\approx5\times10^{52}$~cm$^{3}$, which results in

\begin{equation}
    M_\mathrm{neb} \simeq 3.9 \times10^{-4} (\varepsilon F_\mathrm{H\alpha})^{1/2} \;\; \mathrm{M}_\odot,
\end{equation}

\noindent where $F_{{\mathrm H}\alpha}$ is in units of $10^{-13}$ erg~cm$^{-2}$~s$^{-1}$. 
Thus, the total mass of the nebula around TYC\,3315 is  $0.0012 \times \epsilon^{1/2} \; \mathrm{M}_\odot$. 
Since this nebula does not show limb-brightened emission $\epsilon$ is not small. However, the emission does not seem to fill the nebula completely nor does it seem close to unity. Values between $0.3 \leq \epsilon \leq 0.8$ have been adopted, resulting in a nebular mass in the range $(8\pm2)\times10^{-4}$ M$_\odot$.

\subsection{Complementary multi-wavelength detections}

Wide-field Infrared Survey Explorer ({\it WISE}) images of TYC\,3315 downloaded from the {\it WISE All-Sky Data Release} reveal an emission excess around it in the W1 3.4 $\mu$m, W3 12 $\mu$m, and W4 22 $\mu$m bands (Fig.~\ref{wise}). 

The image in the W1 band, like in H$\alpha$, shows the presence of a parabolic-shaped nebula east of TYC\,3315 (left panel of Fig.~\ref{wise}). At this wavelength the nebula looks wider (up to $\approx$60 arcsec from TYC\,3315) and peaks farther away (at $\approx$45 arcsec from TYC\,3315) than that in H$\alpha$.
This emission is not present in the other {\it WISE} bands. 
Instead, diffuse complex emission is detected in the W3  (right panel of Fig.~\ref{wise}) and W4 bands (not shown here), mostly along the Northeast and Northwest directions.  

The relative spatial distribution of the outer \emph{WISE} W1 and inner H$\alpha$ emissions is observed in various types of nebulae, such as bow-shocks associated with AGB stars \citep[e.g.,][]{Cox2012,OG2023} or Wolf-Rayet wind-blown bubbles \citep{Toala+2015}. 
It is usually interpreted as the stratification of molecular/dust material enveloping an ionized component. However, the prevalence of the arc-like emission in the mid-IR W1 band discards its origin from dust, which would be brighter at longer wavelengths. 
A molecular component, or most likely the bremsstrahlung continuum, like in photoionized regions, might be responsible for this emission \citep{PRL2008,Anderson2012}.

The nature of the emission at longer wavelengths is ambiguous, but it seems to be unrelated to the parabolic nebula around TYC\,3315. 
It might be rather associated with the large-scale diffuse emission across the field of view also seen in the TST H$\alpha$+[N~{\sc ii}] and W1 images. 
This emission could arise from warm dust in the interstellar medium (ISM). 

\begin{figure*}
\begin{center}
\includegraphics[width=0.32\textwidth]{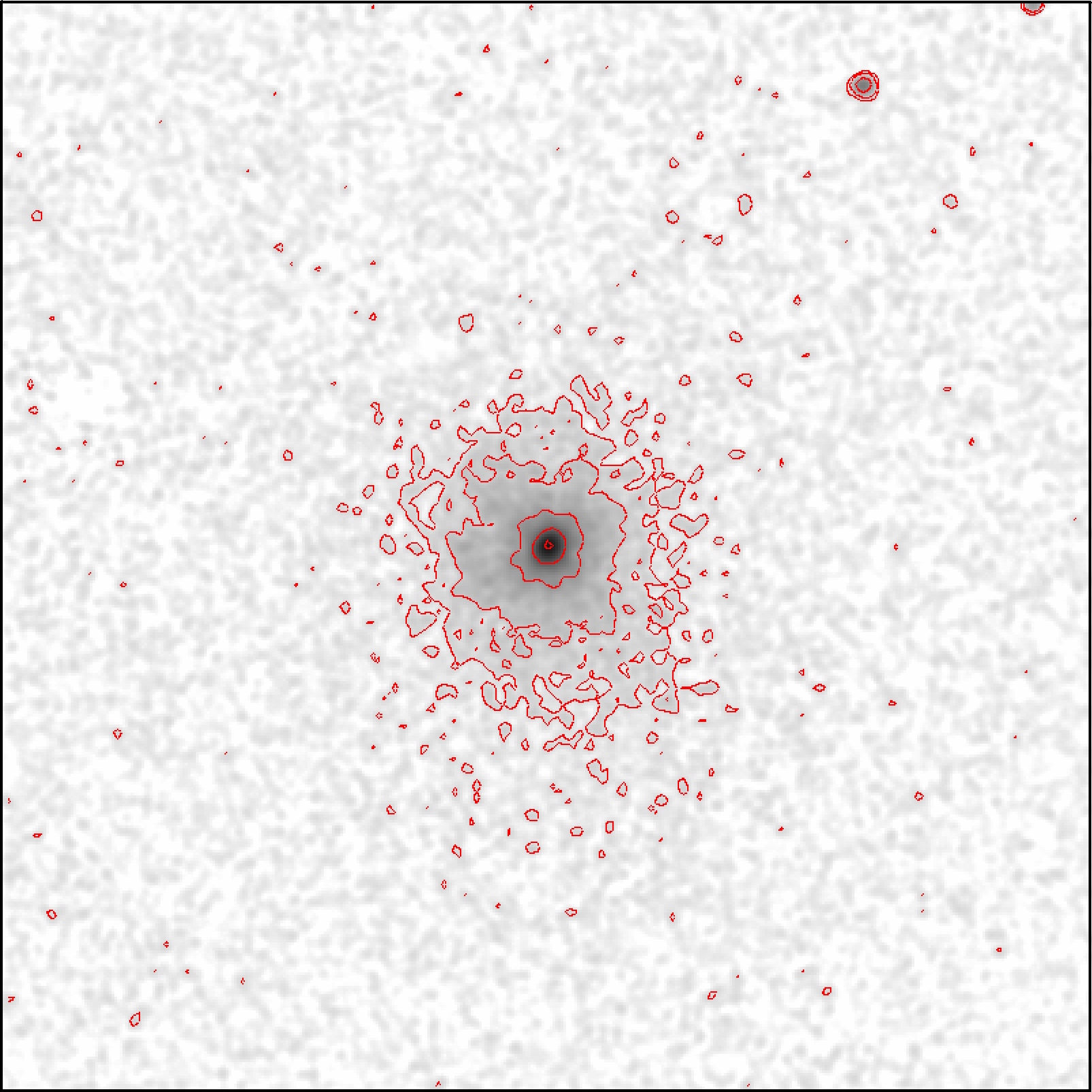} 
\includegraphics[width=0.32\textwidth]{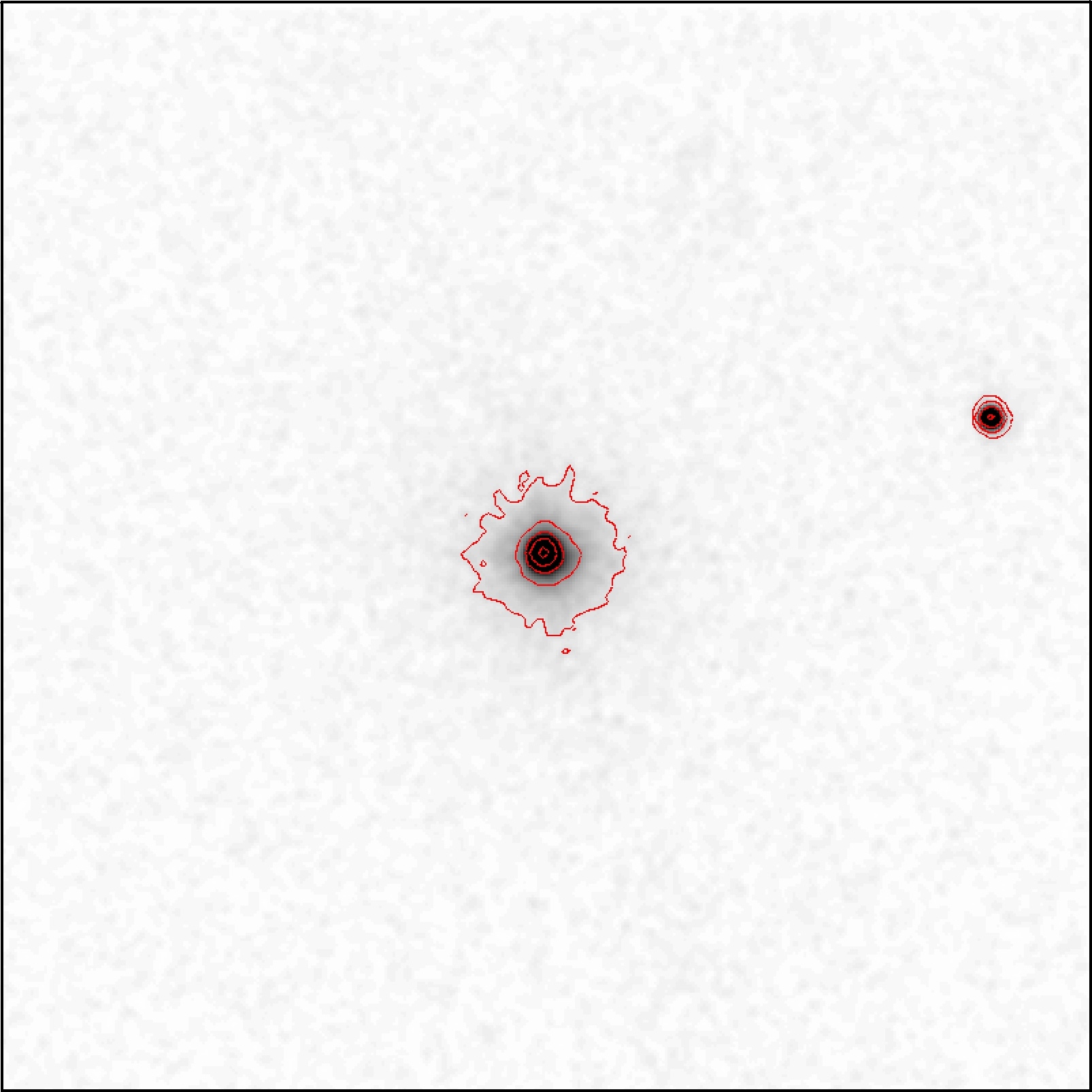}
\includegraphics[width=0.32\textwidth]{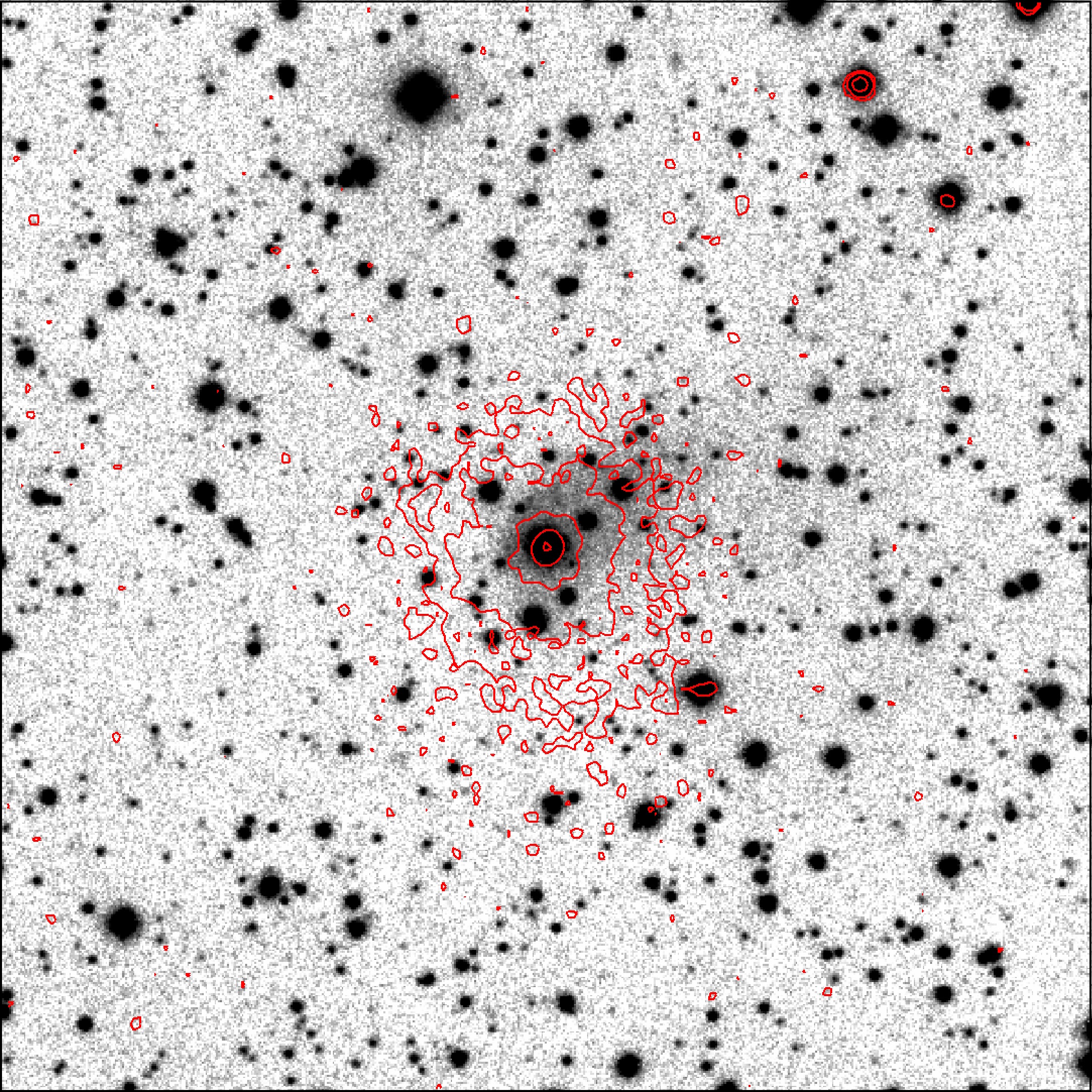}
\end{center}
\caption{
Far-UV {\it GALEX} colour-inverted gray-scale images with red isophotes of TYC\,3315 (left) and a PSF image built from four point-sources in the FoV with [FUV] magnitudes similar to TYC\,3315 (centre), and TST H$\alpha$+[N~{\sc ii}] optical colour-inverted gray-scale image with [FUV] isophotes (right). 
The contours on the [FUV] left and centre panels, which are shown at the same level, probe the presence of diffuse far-UV emission around TYC\,3315. 
Like in Fig.~\ref{wise}, the FoV is $10^\prime\times10^\prime$, and north is up, east to the left.  
}
\label{galex}
\end{figure*}

\begin{figure}
\begin{center}
  \includegraphics[width=0.95\columnwidth]{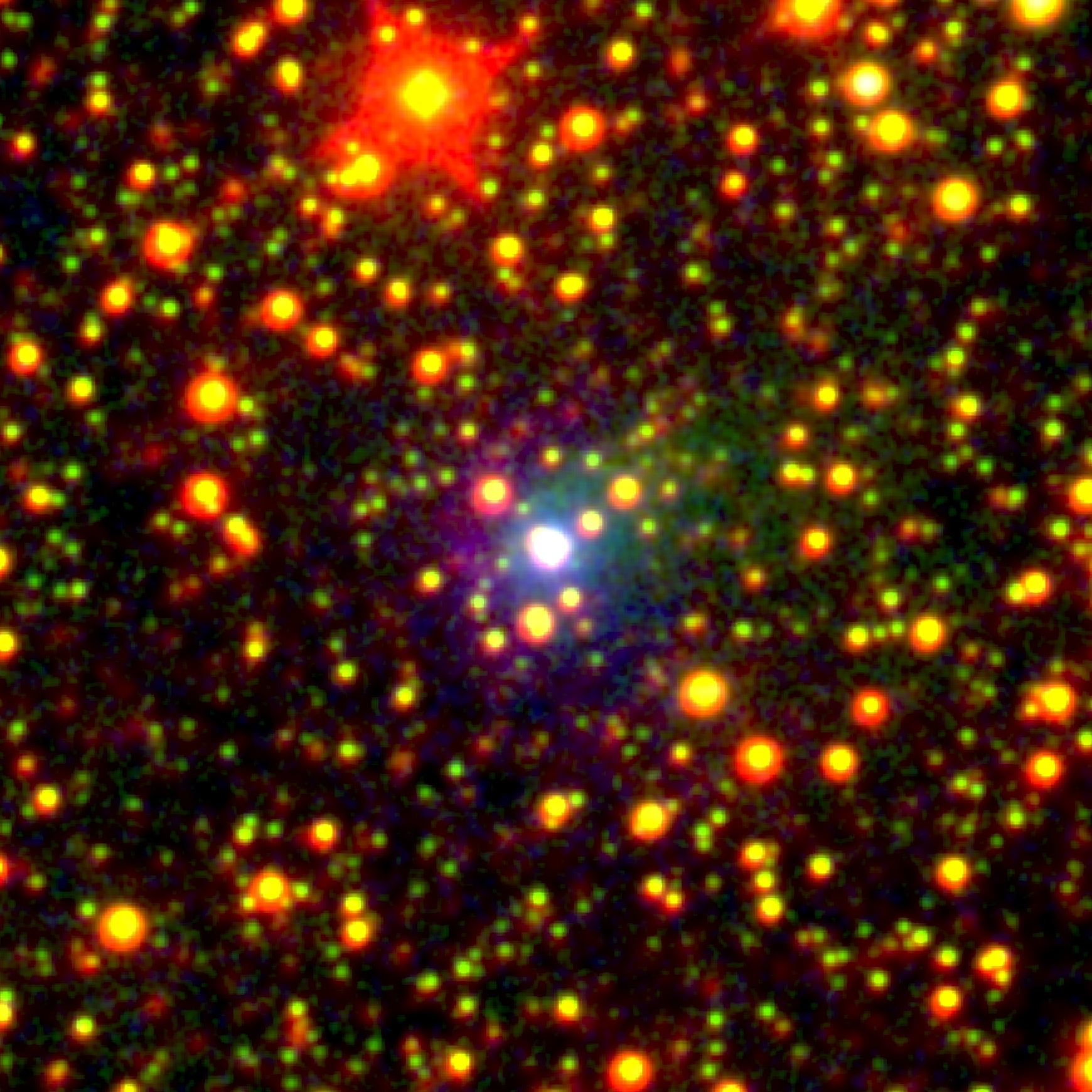}
\end{center}
\caption{
RGB composite colour picture-centred of TYC\,3315 with WISE W1 in red, TST H$\alpha$+[N~{\sc ii}] in green, and GALEX FUV in blue.  
Like in Fig.~\ref{wise}, the FoV is $10^\prime\times10^\prime$, and north is up, east to the left.  
}
\label{fig:rgb}
\end{figure}

The {\it GALEX AIS} far-UV (FUV) observation (tile number \#50055, with exposure time of 96 s) seems to show diffuse emission associated with TYC\,3315 as well (Fig.~\ref{galex}-left). A comparison with the [FUV] stellar PSF derived from stars in the FoV with similar magnitude (Fig.~\ref{galex}-centre) confirms the extended nature of the emission around TYC\,3315. 
On the other hand, no diffuse emission is detected in the {\it GALEX AIS} near-UV ([NUV]) observations of the same field (exposure time 183 s).

A direct comparison between the optical and far-UV emissions from TYC\,3315 reveals that the latter is much more symmetrical and apparently more extended (Fig.~\ref{galex}-right). 
The comparison between IR, optical and far-UV emissions from TYC\,3315 is further illustrated in the RGB colour-composite picture in Fig.~\ref{fig:rgb}.  
The IR emission is wider and more extended than the H$\alpha$, but the H$\alpha$ emission extends farther towards the west direction.  
Whilst the IR and H$\alpha$ emissions have a parabolic-shaped morphology, the far-UV emission looks rather symmetric and less extended around TYC\,3315. Thus, the morphological differences at distinct wavelengths of the diffuse emission around TYC\,3315 indicate a noticeable stratification.

Finally, TYC\,3315 was serendipitously observed by {\it Chandra} on December 20, 2007 (Obs.\ ID 7430, PI G.\ Hussain) for 63.4 ks.  
The star and its nebula are registered on the Front Illuminated CCD S2 of the Advanced CCD Imaging Spectrometer (ACIS).  
A visual inspection shows neither an X-ray counterpart of TYC\,3315 nor its associated bow shock nebula. 
The absence of diffuse X-ray emission is not surprising, since it is often detected in supernova remnants \citep{ChuMacLow1990} and cataclysmic variables \citep[CV,][]{Mukai2003}, but not commonly in nebulae blown by steady mass-loss objects, such as the powerful stellar winds of WR stars \citep{Toala2012,TG2013}, for example.

\begin{table*}
\caption{TYC\,3315-1807-1 and previously known sdB stars associated with nebulae including PHL\,932 \citep{Frew+2010} and those listed in the catalogue compiled by \citet{G-S2021}. 
The apparent sizes and information on morphology were extracted from the {\it HASH} database of PNe \citep{Parker+2016}. 
$z$ is the height over the Galactic plane; the tangential velocity (km s$^{-1}$) has been calculated from the stellar proper motion ($\mu$, mas yr$^{-1}$) and parallax ($p$, mas) using $V_t = 4.741 \times (\mu /p)$.
The nebula Pa\,162, originally in the list of \citet{G-S2021}, is not included below because it has been later noted in the {\it HASH} database to be an artifact on photographic plates.
}
\begin{center}
\begin{tabular}{clcccccc}
\hline
PN\,G name & \multicolumn{1}{c}{Name} & Morphology & Apparent size & Parallax distance & $z$ & $V_r$ & $V_t$ \\
 & & & (arcmin) & (kpc) & (kpc) & (km s$^{-1}$) & (km s$^{-1}$) \\
\hline
$\dots$ & TYC\,3315-1807-1        & Parabolic shaped & $3.5 \times 2.7$   & $0.263 \pm 0.002$ & $0.0371 \pm 0.0003$ & 69.6 & 75.6 \\
\hline
089.1+08.7 & Pa\,39, KPD\,2024$+$5303 & Elliptical     & $0.5 \times 0.4$   & $6.2 \pm 0.8$ & $0.94 \pm 0.12$ & $\dots$ & 198.9 \\
125.9$-$47.0 & PHL\,932 & Parabolic shaped? & $4.0 \times 2.5$ & $0.323 \pm 0.005$ & $0.236 \pm 0.004$ & 1.6 & 56.0 \\
137.6$-$30.0 & Fr\,2-22                 & $\dots$        & 1.0?               & $0.73 \pm 0.02$ & $0.37 \pm 0.01$ & 76.3 & 13.7 \\
306.5$-$31.1 & Fr\,1-4, JL 102          & $\dots$        & $3.7 \times 3.0$   & $1.56 \pm 0.10$ & $0.81 \pm 0.05$ & $\dots$ & 60.0 \\
315.7+42.0 & PaHa\,1, EC\,13290$-$1933            & $\dots$        & $13.0 \times 9.0$ & $1.24 \pm 0.06$ & $0.83 \pm 0.04$ & $\dots$ & 74.9 \\
\hline
\end{tabular}
\end{center}
\label{tab:sdBPN}
\end{table*}

\section{Discussion}

The presence of diffuse emission around sdB stars is unexpected.  
We examine next the possible origin of the ionized nebula around TYC\,3315 and assess the presence of nebular emission around other sdB stars. 

\subsection{On the origin of the ionized nebula around TYC\,3315}

The theoretical prediction of the mass loss rate for an sdB star such as TYC\,3315 ($5.0 < \log g < 6.0$) is $\dot{M} < 10^{-10} \, \mathrm{M}_{\odot}$~yr$^{-1}$ \citep{Unglaub2008}. 
Therefore, the presence of a circumstellar nebula composed of matter ejected by TYC\,3315 is surprising, considering the extremely low mass-loss rate. Alternatively, the nebula could be attributed to the photoionization of the surrounding low-density ISM by the star, although the ionizing flux of TYC\,3315 is expected to be low, considering its low luminosity \citep[$\approx11 L_{\odot}$,][]{Devarapalli2022}. On the other hand, the nebular morphology is highly suggestive of an interaction between the fast-moving star and the local ISM, either through its stellar wind or as an interacting nebula (e.g., an evolved PN).  
Indeed, TYC\,3315 is a high proper motion star, with a space velocity of 102.1 km~s$^{-1}$.

In the next sections we discuss these possibilities and compare them with the observed parameters of the nebula.

\subsubsection{A photoionized Str\"omgren nebula}

TYC 3315 is expected to be a low ionizing flux object. 
Assuming a stellar radius of $\approx 0.15 R_{\odot}$ \citep{Devarapalli2022}, effective temperature $T_{\rm eff}= 29.2$ kK \citep{Kawka2010}, and adopting a blackbody stellar spectrum, its ionizing flux is $Q_{\rm H} \simeq 4 \times 10^{44}$ photon~s$^{-1}$, much lower than a typical PN nucleus ($10^{45} - 10^{48}$ photon~s$^{-1}$). 
However, even relatively low $Q_{\rm H}$ objects can ionize large volumes of the ISM if the gas density is low because the mean free path of ionizing photons increases proportionally to $n_{\rm H}^{-1}$. Then, for the ionizing flux previously determined, the H~{\sc i} Str\"omgren radius ($R_S$) of TYC\,3315 can be estimated by the formula:

\begin{equation}
R_S = \left(\frac{3Q_{\rm H}}{4\pi {\alpha}_B}\right)^{1/3} n_{\rm H}^{-2/3} = 2.34 \, n_{\rm H}^{-2/3},
\end{equation}

\noindent where $n_{\rm H}$ is given in cm$^{-3}$ and $R_S$ in parsecs. Thus, adopting the density of the local ISM $n_{\rm H}=0.2$ cm$^{-3}$ \citep{FS2003,Misiriotis2006,GJ2017} 
TYC\,3315 could ionize hydrogen atoms within a spherical volume of $R_S \simeq 6.8$ pc. At the distance of TYC\,3315, the angular radius of this Str\"omgren sphere would be $\approx 1.5^\circ$, much larger than the nebula detected around TYC 3315. In this case, the low surface-brightness and large angular extent of such a nebula would make its detection very difficult.

Alternatively, the nebula could appear much smaller and brighter if TYC\,3315 had undergone a transient episode of heavy mass-loss increasing the density of the surrounding medium or if the star was moving across an over-density region of the ISM \citep[as suggested in the case of PHL\,932,][]{Frew+2010}. 
Thus, assuming the Str\"omgren radius $R_S$ corresponding to the semi-major extent of the nebula ($\approx 0.13$ pc) the density would be estimated as $n_{\rm H}\simeq 70$ cm$^{-3}$. 
The origin of a transient mass-loss episode in an sdB star would be puzzling, although it has been formerly claimed to explain a common-envelope phase initiated by a low-mass companion \citep[e.g.,][]{Hall+2013}. 

\subsubsection{A fossil planetary nebula or another nebula interacting with the ISM}

The singular morphology of the nebula detected in H$\alpha$ resembles other PNe distorted by interaction with the ISM.  
These are most likely to be found in high density environments \citep{Borkowski1990}, but large, evolved PNe with low densities and large ``cross sections'' can experience similar effects if moving fast enough through a low density ISM \citep{TK1996}. 
However, the location of TYC \,3315 on the HR diagram, far below the cooling track of WDs, implies a long time since it ejected (if ever) a PN \citep{MB2016}. 
Therefore, the PN is expected to have dispersed into the ISM a long time ago.  
Some objects may mimic PNe, but have a different origin. 

One such example is Abell\,35, a low surface brightness nebula around a star travelling at the speed of 150 km~s$^{-1}$ through the ISM. 
The detailed morphology of this nebula is indeed indicative of interaction with the ISM \citep{Hollis+1996}, with H$\alpha$ images showing a bow-shock nebula along the direction of the star's motion, embedded within a large round nebula, whereas the higher excitation [O~{\sc iii}] emission lines delineate a smaller, round nebula \citep{Ziegler+2012}.
The central star of Abell\,35, a well-known binary system with a G8 III-IV-type companion, has an estimated mass-loss rate of $\dot{M} = 3 \times 10^{-9} M_{\odot}$ yr$^{-1}$ \citep{Jacoby1981}, i.e., at least 30 times higher than that of TYC\,3315.
Otherwise the PN nature of Abell\,35 has been questioned, suggesting that it is a bow-shock nebula surrounded by a photoionized Str\"ongren sphere in the ISM \citep{FP2010}. 

Other sources with some morphological resemblance are EGB\,4 \citep{EGB1984} and Fr\,2-11 \citep{FMP2006}, two bow-shock shaped nebulae associated with the CV systems BZ\,Cam and V341\,Ara, respectively.
These are CV systems of the UX\,UMa subclass that exhibit high accretion rates onto the WD from their main-sequence donors.
These nebulae have been suggested to arise from the strong stellar wind of BZ\,Cam, as it moves supersonically through the ISM \citep{Hollis+1992}, and from the high-speed encounter of V341\,Ara with a small ISM cloud or with its own ejecta from a past nova outburst \citep{BM2018}. 
Their ionized masses could be just a few times larger than that of the nebula around TYC\,3315. The excitation mechanism in both systems, and in TYC 3315 also, may be a mix of shocking and photoionization, with shocks prevailing at the bow shock location and photoionization at work in the most extended regions.

\subsubsection{A bow shock nebula}

The interaction of the stellar wind with the local ISM as the star travels at a supersonic speed can result in the formation of a bow-shock nebula. 
This could be the case of TYC\,3315, whose space velocity is $V_{\star}=102.1$ km~s$^{-1}$. 
The shock between the stellar wind and the gas in the ISM is expected to occur along the direction of the star's motion on the plane of the sky (PA $\simeq98^\circ$). 
This value is in reasonable agreement with the orientation of the apex of the parabolic nebula around TYC\,3315: $\approx 92^\circ$.  

The pressure balance in a bow shock formed by the interaction between the stellar wind with density $\rho_{\rm w}$ and terminal wind velocity $V_{\rm w}$ of a star moving with velocity $V_\star$ and the local ISM of density $\rho_{\rm ISM}$

\begin{equation}
    \rho_{\rm w} \, V_{\rm w}^2 = \rho_{\rm ISM} \, V_\star^2 
\end{equation}
can be rewritten as \citep{BKK1971,Wilkin1996}:

\begin{equation}
\dot{M} \, V_{\rm w} = 4\pi \, {\rho}_{\rm  ISM} \, R_0^2 \, V_{\star}^2 \;\;\; [{\rm g \; cm}~{\rm s}^{-2}], 
\label{equation1}
\end{equation}

\noindent where $\dot{M}$ is the stellar mass-loss rate and $R_0$ is the bow shock distance to the star. Using conventional units, the equation above can be rewritten as

\begin{equation}
\begin{split}
\dot{M}V_{\rm w} \; [(\mathrm{M}_{\odot}\,\mathrm{yr}^{-1})\;(\mathrm{km}\,\mathrm{s}^{-1})] = 1.90 \times 10^{17} \; {\rho}_{\rm  ISM} \, [{\rm g \;cm}^{-3}] \\ \; R_0^2 \, [{\rm pc}]^2 \; V_{\star}^2 
 \, [{\rm km \; s}^{-1}]^2.
 \end{split}
\label{momentum}
\end{equation}

Assuming a fractional abundance [He/H]=0.1 \citep{GL1992,Kilian1992} and the local ISM density of 0.2 H atoms cm$^{-3}$, the local ISM gas density would be ${\rho}_{\rm ISM} = 5 \times 10^{-25}$ g cm$^{-3}$. At a distance of 263.1 pc, the $\approx28^{\prime\prime}$ separation between the star and the apex of the parabolic shaped nebula corresponds to $R_{\rm 0,proj} \simeq 0.036$ pc. 
Assuming that this is aligned with the motion of the star with an inclination relative to the plane of the sky of $\approx 43^\circ$, we have $R_0 \simeq 1.37 \times R_{\rm 0,proj} \simeq 0.049$ pc. The equation above then implies 

\begin{equation}
\dot{M} \; V_{\rm w} \simeq 2.3 \times 10^{-6} \; [(\mathrm{M}_{\odot}~\mathrm{yr}^{-1})~(\mathrm{km~s}^{-1})]. 
\label{momentum2}
\end{equation}

The terminal wind velocity can be expected to be comparable or higher than the escape velocity, which is $\simeq 1000$ km~s$^{-1}$ for a star of $0.57~\mathrm{M}_{\odot}$ and a radius of $0.217~\mathrm{R}_{\odot}$ \citep{Schaffenroth2022} or $\approx 835$ km~s$^{-1}$ for a star of $0.274~\mathrm{M}_{\odot}$ and a radius of $0.150~\mathrm{R}_{\odot}$ \citep{Devarapalli2022}, implying $\dot{M} \lesssim [2.3-2.8]\times10^{-9}~\mathrm{M}_{\odot}$~yr$^{-1}$. However, this mass-loss rate is 20 to 30 times higher than the theoretical prediction for an sdB star with $5.0 < \log g < 6.0$ \citep[$\dot{M} < 10^{-10} \, \mathrm{M}_{\odot}$~yr$^{-1}$,][]{Unglaub2008}. 

This approach assumes that the stellar wind is fully driven by the stellar radiation pressure. If $\eta$ is the efficiency of this mechanism the momentum transferred by radiation to the stellar wind can be related to the stellar luminosity using the relation

\begin{equation}
\frac{L_{\star}}{c} \eta = \dot{M} \; V_{\rm w}, 
\label{luminosity}
\end{equation}

\noindent i.e., the stellar wind momentum is proportional to the stellar luminosity. 
Substituting the result obtained by equation \ref{momentum2} into equation \ref{luminosity}, and assuming $0.1 < \eta < 1$, we obtain $L_{\star} \simeq 10^3 - 10^4$~L$_{\odot}$. However, the luminosity of TYC\,3315 is much lower \citep[$\approx$11 L$_\odot$,][]{Devarapalli2022}, leading to the conclusion that the radiation pressure of TYC\,3315 alone cannot provide the mechanical energy required to produce the observed bow shock.

In order to assess the origin of the bow shock it is necessary to determine the relative contributions to the mass of the swept ISM and the TYC\,3315 stellar wind.  
For the ISM density of 5 $\times 10^{-25}$ g cm$^{-3}$ described above, the swept mass of the ISM would be 1.2 $\times10^{-5}$ M$_{\odot}$.  
As for the stellar wind, it can be assumed that it contributed an amount of mass to the bow shock for a period of time similar to its age.  
The proper motion of the star, $\mu = 59.287\pm0.043$ mas~yr$^{-1}$, requires a time-lapse $\approx$3500 yr to move along the $\approx$3.5 arcmin long nebula.  
Thus, the mass-loss rate of $\dot{M} < 10^{-10} M_{\odot}~\mathrm{yr}^{-1}$ predicted by the models \citep{Unglaub2008} would have been able to assemble only $<2\times10^{-7}~\mathrm{M}_\odot$ along the direction of the star's motion.
Therefore, the joint contributions of the swept ISM and stellar wind constitute only a small fraction of the nebular mass, which has been estimated to be $(8\pm2) \times 10^{-4}$ M$_\odot$ (Sect. \ref{halpha}).

\subsubsection{Concluding remarks}

All of the above considerations make the presence of an ionized nebula around TYC \,3315 even more intriguing. A Str\"omgren sphere in the low density ISM is unreasonable, unless the star had experienced a previous episode of mass loss of unknown origin or it was by chance crossing an over-density region of the ISM. 
An evolved PN interacting with the ISM seems to be out of question, as such a nebula would have dispersed long ago, given the time scale from the post-AGB phase to the present stage of TYC\,3315. 
Finally, the stellar wind is uncapable of sustaining a bow-shock of the size of the nebula around TYC\,3315.

The case of TYC\,3315 is somehow similar to 1RXS\,J052832.5+28382, a CV system where a bow shock nebula has been detected very recently \citep[][]{Ilkiewicz2026}. 
1RXS\,J052832.5+28382 has a short orbital period of only 80 minutes and a strong magnetic field ($B \simeq 42 - 45$ MG), which conforms with its classification as a polar CV. 
More importantly, it lacks an accretion disc that could power a disc wind. 
Both TYC\,3315 and 1RXS\,J052832.5+28382 move at high speed ($V_{\star}=102$ and 142 km s$^{-1}$, respectively), but their stellar luminosities are too low and their stellar winds too weak to form a bow shock with the observed dimensions.

It is important to note that the nebula around TYC\,3315 does not appear detached from the star as commonly observed in bow-shock nebulae \citep[e.g.,][]{Toala+2016}. 
On the other hand, the coincidence between the apex of the parabolic-shaped nebula and the direction of star's proper motion on the plane of the sky makes reasonable to attribute the nebula to the interaction of TYC\,3315 with its surrounding ISM. 
The nebula around TYC\,3315 could rather be a Mach wake in the ISM, where the stellar wind is unable to compress the gas ahead
into a thin layer to form a bow-shock, but it still can produce an over-density in the ISM. 
Still, the recombination time scales would be long enough to allow the survival of an ionized nebula a few thousand years after TYC\,3315 has passed through it. 

\subsection{Other sdB stars associated with nebulae or shells}

The sample of PNe in Gaia EDR3 compiled by \citet{G-S2021} includes 5 nebulae with an sdB central star in the HASH Database \citep{Parker+2016,Bojicic2017}, namely Fr\,1-4, Fr\,2-22, PaHa\,1, Pa\,39, and Pa\,162. 
There is also a long-known nebula around the sdB star PHL\,932 \citep{AS1967}. 
The ``nebula'' Pa\,162 has recently been noted to be a plate artifact in the HASH Database of PNe and it is thus been disregarded for further discussion. 
The available information on the other five sources is listed in Table~\ref{tab:sdBPN}. 

The nature and authenticity of these nebulae are questionable. 
Fr\,1-4, Fr\,2-22, and PaHa\,1 are classified as possible PNe in the HASH Database, whereas Pa\,39 is considered a likely PN. 
The PN nature of Fr\,2-22 has been further questioned by \citet{Hillwig+2022}, even though just on the basis of the membership of the sdB star (GALEX\,J015054.4$+$310745) to a binary system with a He WD companion. In addition, the nebular source is undetected (Guerrero et al., in prep.) in large FoV H$\alpha$+[N~{\sc ii}] images from the Javalambre Photometric Local Universe Survey (J-PLUS) DR3 obtained with the 80-cm Javalambre Auxiliary Survey Telescope (JAST80) at the Observatorio Astrofísico de Javalambre ({\it OAJ}).

Morphological information about the nebulae in Table ~\ref{tab:sdBPN} is scarce. 
Pa\,39 (PN\,G089.1+08.7) is described in the HASH Database as an oval-shaped nebula, whilst  PHL\,932 (PN\,G125.9$-$47.0) shows many morphological similarities with the nebula around TYC\,3315. These are probably the only three bona fide nebulae associated with sdB stars known to date.
The interaction between the central star of Pa\,39 (KPD\,2024$+$5303) and the ISM cannot be ascertained because its galactic velocity cannot be well determined due to its large distance (6.21 kpc) and its position, outside the galactic solar circle ($l = 89^{\circ}$).
On the other hand, the nebula around PHL\,932 could have been formed because of its relatively high speed. 
This nebula, originally interpreted as a PN \citep{Mendez+1988}, was later described as a small Str\"omgren sphere in a region of over-density of the ISM \citep{Frew+2010}. 

The high speed of an sdB star can certainly favor the formation of an associated nebula in the form of a bow shock or a wake between its stellar wind and the local ISM.
There are other conditions necessary to form an ionized nebula, such as high mass-loss rate, stellar luminosity and density of the ISM. 
Apart from the central star of Pa\,39, all the other sdBs in Table ~\ref{tab:sdBPN} can be reliably classified as ``runaway stars'', since their velocity exceeds 30 km~s$^{-1}$ \citep{Torres2025}. The density of the ISM tends to be higher in regions closer to the Galactic plane, which is the case for TYC\,3315 (and to some extent PHL\,932), but not for Pa\,39. 
These are trends that have been observed among other classes of bow shocks such as those around red supergiants, AGB stars \citep{Cox2012,GO2023}, and runaway young massive stars \citep{Peri2015,Kobulnicky2016}.
Deep narrow-band imaging of these sources will assess whether these sdB stars are really surrounded by ionized nebulae.

\section{Conclusions}

The number of nebulae or shells around sdB stars is very small: 
only Pa\,39 around KPD\,2024$+$5303 and those reported around PHL\,932 and TYC\,3315 seem to be bona fide nebular sources. Thus, it is imperative to confirm the detection of nebulae around the sources listed in Table~\ref{tab:sdBPN} and obtain an accurate description of their morphologies. 

Otherwise, the low ionizing flux and the weak wind momentum provided by the stellar radiation of sdB stars could be the reason for the small number of nebulae around them.  
Therefore, the detection of a parabolic shaped nebula in front of TYC\,3315 requires additional mechanisms to enhance the mass-loss rate predicted by stellar models, which is clearly insufficient to form and keep this bow shock. This is also the case for the discless polar CV 1RXS\,J052832.5+28382. 
Perhaps the cool low-mass companion of TYC\,3315 plays a role to explain this discrepancy, having contributed to the formation of a post-common-envelope nebula.

\section*{acknowledgments}
{We thank the referee Dr.\ Albert Zijlstra for his valuable suggestions that greatly improved the quality of the manuscript and Dr.\ Lida Oskinova for the valuable discussion.
R.O.\ thanks the support of the São Paulo Research Foundation (FAPESP), grants \#2023/05298-0 and \#2025/11753-7. 
F.B.\ and M.A.G.\ acknowledge financial support from grants CEX2021-001131-S funded by MCIN/AEI/10.13039/501100011033, PID2022-142925NB-I00 from the Spanish Ministerio de Ciencia, Innovaci\'on y Universidades (MCIU) cofunded with FEDER funds.
M.A.\ acknowledges financial support from the MCIU grant PID2024-162229NB-I00 cofunded with FEDER funds.  
This article is based on observations made in the Transient Survey Telescope (TST\footnote{\url{http://ttt.iac.es}}) sited at the Teide Observatory of the Instituto de Astrofísica de Canarias that Light Bridges operates in the island of Tenerife, Canary Islands (Spain) and the Observatorio de
Sierra Nevada (OSN), operated by the Instituto de Astrofísica de Andalucía (IAA-CSIC).
The observation time rights (DTO) used for this research on TST were consumed in the PEI "SBSTLLAR25". This research used storage and computing capacity in ASTRO POC's EDGE computing center at Tenerife under the form of Indefeasible Computer Rights (ICR), consumed in the PEI "SBSTLLAR25". The ICRs used for this research were provided by Light Bridges in cooperation with Bechtle AG. Dr. Antonio Maudes’s insights in economics and law were instrumental in shaping the development of this work.
We thank the OSN technicians and the telescope operator, Victor Casanova, for their support, and Diana Korotun for providing us with the gaia-spectra script\footnote{\href{https://archive.softwareheritage.org/swh:1:cnt:64db8bc6600e3142ccb6d099b22efcbe7274374c;origin=https://gitlab.com/makhlaghi/useful-scripts.git;visit=swh:1:snp:51b6b0619cead16eb70c6ce11c5436e99f99ab60;anchor=swh:1:rev:40da995a5d5d7094b40f176e7a176bf407ce889f;path=/gaia-spectra.sh}{\texttt{swh:1:cnt:64db8bc6600e3142ccb6d099b22efcbe7274374c}}} used in the flux calibration of the images.
This research has made extensive use of NASA’s Astrophysics Data System and the SIMBAD database, operated at CDS, Strasbourg, France, and HASH, an online database at the Laboratory for Space Research at HKU that federates available multi-wavelength imaging, spectroscopic, and other data for all known Galactic PNe and is available at: http://www.hashpn.space.


\section{Data availability}

Some of the data underlying this article are publicy available in: 
the {\it GALEX tile search}, hosted by {\it The Barbara Mizulski Archive for Space Telescopes}, at the {\it GALEX} website {\it http://galex.stsci.edu/gr6/?page=tilelist\&survey=allsurveys}; the {\it WISE} all-sky survey hosted by the {\it NASA/IPAC Infrared Science Archive} available at  {\it irsa.ipac.caltech.edu/Missions/wise.html}; the {\it GAIA DR3 catalogue}, hosted by the {\it GAIA ESA Archive} at {\it gea.esac.esa.int/archive}.

\bsp    
\label{lastpage}


\begin{thebibliography}{99}

\bibitem[\protect\citeauthoryear{Akhlaghi}{2019a}]{gnuastroSegment2019a}Akhlaghi M., 2019a, preprint (arXiv:1909.11230), https://doi.org/10.48550/arXiv.1909.11230

\bibitem[\protect\citeauthoryear{Akhlaghi}{2019b}]{makecatalog2019b}
Akhlaghi M., 2019b, ASP Conf. Ser., Vol. 521, Separating Detection and Catalog Production. Astron. Soc. Pac. San Francisco, p.299

\bibitem[\protect\citeauthoryear{Akhlaghi}{2025}]{gnuastro2025} 
Akhlaghi M., 2025, GNU Astronomy Utilities 0.24, Version 0.24, Free Software Foundation, doi:10.5281/zenodo.17726900

\bibitem[\protect\citeauthoryear{Akhlaghi \& Ichikawa}{2015}]{gnuastro2015}Akhlaghi M. \& Ichikawa T., 2015, \apjs, 220, 1

\bibitem[\protect\citeauthoryear{Aller}{2015}]{Aller2015} Aller, A., 2015, PhD Thesis, Departamento de Física Aplicada, Universidade de Vigo, Vigo, Spain

\bibitem[\protect\citeauthoryear{Aller et al.}{2015}]{Aller+2015}
Aller, A., Miranda, L.F., Olgu\'\i n, L., V\'azquez, R., Guill\'en, P.F., Oreiro, R., Ulla, A. \& Solano, E., 2015\, \mnras, 446, 317

\bibitem[\protect\citeauthoryear{Anderson et al.}{2012}]{Anderson2012}Anderson, L.D., Zavagno, A., Barlow, M.J., Garc\'\i a-Lario, P. \& Noriega-Crespo, A., 2012, \aap, 537, A1

\bibitem[\protect\citeauthoryear{Arp \& Scargle}{1967}]{AS1967} 
Arp H., Scargle J.~D., 1967, ApJ, 150, 707. 

\bibitem[\protect\citeauthoryear{Baranov, Krasnobaev, \& Kulikovskii}{1971}]{BKK1971} 
Baranov V.~B., Krasnobaev K.~V., Kulikovskii A.~G., 1971, SPhD, 15, 791


\bibitem[\protect\citeauthoryear{Boji\v{c}i\'c et al.}{2017}]{Bojicic2017}
Boji\v{c}i\'c, I. S., Parker, Q. A., \& Frew, D. J., 2017, in {\it Planetary Nebulae: Multi-Wavelength Probes of Stellar and Galactic Evolution}, 323, eds. X. Liu, L. Stanghellini, \& A. Karakas, p. 327

\bibitem[\protect\citeauthoryear{Bond \& Miszalski}{2018}]{BM2018} 
Bond H.~E., Miszalski B., 2018, PASP, 130, 094201. 

\bibitem[\protect\citeauthoryear{Borkowski et al.}{1990}]{Borkowski1990}Borkowski, K.J., Sarazin, C,L. \& Soker, N., 1990, \apj, 360, 173

\bibitem[\protect\citeauthoryear{Boyarchuk et al.}{1968}]{Boyarchuk+1968} 
Boyarchuk A.~A., Gershberg R.~E., Godovnikov N.~V., Pronik V.~I., 1968, IAUS, 34, 162

\bibitem[\protect\citeauthoryear{Chu \& Mac Low}{1990}]{ChuMacLow1990}Chu, Y.-H. \& Mac Low, M.-M., 1990, \apj, 365, 510

\bibitem[\protect\citeauthoryear{Cox et al.}{2012}]{Cox2012}Cox, N.L.J., Kerschbaum, F., van Marle, A.-J., Decin, L., Ladjal, D., Mayer, A., Groenewegen, M.A.T., van Eck, Royer, P., Ottensamer, R. et al., \aap, 537, A35

\bibitem[\protect\citeauthoryear{Devarapalli et al.}{2022}]{Devarapalli2022}Devarapalli, S.P., Jagirdar, R., Gundeboina, V.K., Thomas, V.S. \& Mynampati, S.R., 2022, AJ, 164, 11

\bibitem[\protect\citeauthoryear{Ellis, Grayson, \& Bond}{1984}]{EGB1984} 
Ellis G.~L., Grayson E.~T., Bond H.~E., 1984, PASP, 96, 283. 

\bibitem[\protect\citeauthoryear{Eskandarlou et al.}{2023}]{Eskandarlou2023}Eskandarlou S., Akhlaghi M., Infante-Sainz R., Saremi E., Raji S., Sharbaf Z., Golini G., Ghaffari Z., Knapen J.~H., 2023, Res. Notes AAS, 7, 269. doi:10.3847/2515-5172/ad14f4

\bibitem[\protect\citeauthoryear{Frew et al.}{2010}]{Frew+2010} 
Frew D.~J., Madsen G.~J., O'Toole S.~J., Parker Q.~A., 2010, PASA, 27, 203. 

\bibitem[\protect\citeauthoryear{Frew, Madsen, \& Parker}{2006}]{FMP2006} 
Frew D.~J., Madsen G.~J., Parker Q.~A., 2006, IAUS, 234, 395. 

\bibitem[\protect\citeauthoryear{Frew \& Parker}{2010}]{FP2010} 
Frew D.~J., Parker Q.~A., 2010, PASA, 27, 129. 

\bibitem[\protect\citeauthoryear{Frisch \& Slavin}{2003}]{FS2003}Frisch, P.C. \& Slavin, J.D., 2003, \apj, 594, 844

\bibitem[\protect\citeauthoryear{Gaia collaboration}{2023}]{Gaia2023}Gaia Collaboration, 2023, \aap, 674, 1



\bibitem[\protect\citeauthoryear{Gies \& Lambert}{1992}]{GL1992}Gies, D.R. \& Lambert, D.L., 1992, \apj, 387, 673

\bibitem[\protect\citeauthoryear{Gonz\'alez-Santamar\'\i a et al.}{2021}]{G-S2021}Gonz\'alez-Santamar\'\i a, I., Manteiga, M., Manchado, A., Ulla, A., Dafonte, C. \& L\'opes Varela, P., 2021, \aap, 656, A51


\bibitem[\protect\citeauthoryear{Gry \& Jenkins}{2017}]{GJ2017}Gry, C. \& Jenkins, E.B., 2017, \aap, 598, 31

\bibitem[\protect\citeauthoryear{Guerrero \& Ortiz}{2023}]{GO2023}Guerrero, M.A. \& Ortiz, R., 2023, \mnras, 527, 4730

\bibitem[\protect\citeauthoryear{Hall et al.}{2013}]{Hall+2013} 
Hall P.~D., Tout C.~A., Izzard R.~G., Keller D., 2013, MNRAS, 435, 2048. 

\bibitem[\protect\citeauthoryear{Heber}{2009}]{Heber2009}Heber, U., 2009, ARA\&A, 47, 211

\bibitem[\protect\citeauthoryear{Hillwig et al.}{2022}]{Hillwig+2022} 
Hillwig T.~C., Reindl N., Rotter H.~M., Rengstorf A.~W., Heber U., Irrgang A., 2022, MNRAS, 511, 2033.

\bibitem[\protect\citeauthoryear{Hollis et al.}{1992}]{Hollis+1992} 
Hollis J.~M., Oliversen R.~J., Wagner R.~M., Feibelman W.~A., 1992, ApJ, 393, 217. 

\bibitem[\protect\citeauthoryear{Hollis et al.}{1996}]{Hollis+1996} 
Hollis J.~M., van Buren D., Vogel S.~N., Feibelman W.~A., Jacoby G.~H., Pedelty J.~A., 1996, ApJ, 456, 644. 

\bibitem[\protect\citeauthoryear{Ilkiewicz et al.}{2026}]{Ilkiewicz2026} Ilkiewicz, K., Scaringi, S., de Martino, D., Knigge, C., Motta, S.E., Rea, N., Buckley, D., Castro-Segura, N., Groot, P.J., McLeod, A.F., Parker, L.T. \& Veresvarska, M., 2026, Nature Astron., 10, 391

\bibitem[\protect\citeauthoryear{Jacoby}{1981}]{Jacoby1981}Jacoby, G.H., 1981, ApJ, 244, 903

\bibitem[\protect\citeauthoryear{Kawka et al.}{2010}]{Kawka2010}Kawka, A, Vennes, S., N\'emeth, P., Kraus, M. \& Kub\'at, J., 2010, \mnras, 408, 992

\bibitem[\protect\citeauthoryear{Kilian}{1992}]{Kilian1992}Kilian, J., 1992, \aap, 262, 171

\bibitem[\protect\citeauthoryear{Kobulnicky et al.}{2016}]{Kobulnicky2016}Kobulnicky, H.A., Chick, W.T., Schurhammer, D.P., Andrews, J.E., Povich, M.S., Munari, S.A., Olivier, G.M., Sorber, R.L., Wernke, H.N., Dale, D.A. \& Dixon, D.M., 2016, \apjs, 227, 18

\bibitem[\protect\citeauthoryear{Maxted et al.}{2001}]{Maxted2001}Maxted, P.F.L., Heber, U., Marsh, T.R., North, R.C., 2001, \mnras, 326, 1391




\bibitem[\protect\citeauthoryear{Mendez et al.}{1988}]{Mendez+1988} 
Mendez R.~H., Groth H.~G., Husfeld D., Kudritzki R.~P., Herrero A., 1988, A\&A, 197, L25

\bibitem[\protect\citeauthoryear{Miller Bertolami}{2016}]{MB2016} 
Miller Bertolami M.~M., 2016, A\&A, 588, A25. 

\bibitem[\protect\citeauthoryear{Misiriotis et al.}{2006}]{Misiriotis2006}Misiriotis, A., Xilouris, E. M., Papamastorakis, J., Boumis, P. \& Goudis, C. D., 2006, \aap, 459, 113


\bibitem[\protect\citeauthoryear{Mukai et al.}{2003}]{Mukai2003}Mukai, K., Kinkhabwala, A., Peterson, J.R., Kahn, S.M. \& Paerels, F., 2003, \apj, 586, L77

\bibitem[\protect\citeauthoryear{Napiwotzki et al.}{2004}]{Napiwotzki2004}Napiwotzki, R., Karl, C.A., Lisker, T., Heber, U., Christlieb, N. et al., 2004, Astron. Space Sci., 291, 321


\bibitem[\protect\citeauthoryear{Osterbrock \& Ferland}{2006}]{OF2006} 
Osterbrock D.~E., Ferland G.~J., 2006, Astrophysics of Gaseous Nebulae and Active Galactic Nuclei, second edition, The MIT Press, Cambridge, USA

\bibitem[\protect\citeauthoryear{Parker, Boji{\v{c}}i{\'c}, \& Frew}{2016}]{Parker+2016} 
Parker Q.~A., Boji{\v{c}}i{\'c} I.~S., Frew D.~J., 2016, JPhCS, 728, 032008. 


\bibitem[\protect\citeauthoryear{Peri et al.}{2015}]{Peri2015}Peri, C.S. Benaglia, P. \& Isequilla, N.L., 2015, \aap, 578, A45

\bibitem[\protect\citeauthoryear{Phillips \& Ramos-Larios}{2008}]{PRL2008}Phillips, J.P. \& Ramos-Larios, G., 2008, \mnras, 383, 1029


\bibitem[\protect\citeauthoryear{Sahman et al.}{2015}]{Sahman+2015} 
Sahman D.~I., Dhillon V.~S., Knigge C., Marsh T.~R., 2015, MNRAS, 451, 2863. 

\bibitem[\protect\citeauthoryear{Schaffenroth et al.}{2022}]{Schaffenroth2022}Schaffenroth, V., Pelisoli, I., Barlow, B.N., Geier, S. \& Kupfer, T., 2022, \aap, 666, 182

\bibitem[\protect\citeauthoryear{Stark \& Wade}{2003}]{SW2003}Stark, M.A. \& Wade, R.A., 2003, AJ, 126, 1455

\bibitem[\protect\citeauthoryear{Stroeer et al.}{2007}]{Stroer2007}
Stroeer, A. Heber, U., Lisker, T., Napiwotzki, R., Dreizler, S. et al., 2007, \aap, 462, 269

\bibitem[\protect\citeauthoryear{Toal\'a et al.}{2012}]{Toala2012}
Toal\'a, J.A., Guerrero, M.A., Chu, Y.-H., Gruendl, R.A., Arthur, S.J., Smith, R.C. \& Snowden, S.L., 2012, \apj, 755, 77

\bibitem[\protect\citeauthoryear{Toal\'a \& Guerrero}{2013}]{TG2013}
Toal\'a, J.A. \& Guerrero, M.A., 2013, \aap, 559, 52

\bibitem[\protect\citeauthoryear{Toal{\'a} et al.}{2015}]{Toala+2015} 
Toal{\'a} J.~A., Guerrero M.~A., Ramos-Larios G., Guzm{\'a}n V., 2015, A\&A, 578, A66

\bibitem[\protect\citeauthoryear{Toal{\'a} et al.}{2016}]{Toala+2016} 
Toal{\'a} J.~A., Oskinova L.~M., Gonz{\'a}lez-Gal{\'a}n A., Guerrero M.~A., Ignace R., Pohl M., 2016, \apj, 821, 79. 

\bibitem[\protect\citeauthoryear{Torres et al.}{2025}]{Torres2025}Torres, G., Neuh\"auser, R., H\"uttel, S.A. \& Hambaryan, V.V., 2025, \mnras, 539, 282

\bibitem[\protect\citeauthoryear{Tweedy \& Kwitter}{1996}]{TK1996} 
Tweedy R.~W., Kwitter K.~B., 1996, ApJS, 107, 255. 


\bibitem[\protect\citeauthoryear{Unglaub}{2008}]{Unglaub2008}Unglaub, K., 2008, \aap, 486, 923

\bibitem[\protect\citeauthoryear{Wilkin}{1996}]{Wilkin1996}Wilkin, F.P., 1996, \apj, 459, L31

\bibitem[\protect\citeauthoryear{Wo\'zniak et al.}{2004}]{Wozniak2004}Wo\'zniak, P.R., Vestrand, W.T., Akerlof, C.W., Balsano, R., Bloch, J., Casperson, D., Fletcher, S., Gisler, G., Kehoe, R., Kinemuchi, K. et al., 2004, \aj, 127, 2436

\bibitem[\protect\citeauthoryear{Ziegler et al.}{2012}]{Ziegler+2012} 
Ziegler M., Rauch T., Werner K., K{\"o}ppen J., Kruk J.~W., 2012, \aap, 548, A109. 

\end{thebibliography}
\end{document}